\documentclass[runningheads]{svmult}

\usepackage{makeidx}   
\usepackage{graphicx}  
\usepackage{subeqnar}  
\usepackage{multicol}  
\usepackage{physprbb}  
\makeindex             

\newcommand{\etal}{{\it et~al.\ }}
\newcommand{\ie}{{\it i.e.,}\ }
\newcommand{\arcsec}{{\tt ''}}
\newcommand{\gtrsim}{\ {\hskip-4pt\raise-.5ex\hbox{$\buildrel>\over\sim$}}\ }
\newcommand{\lesssim}{\ {\hskip-4pt\raise-.5ex\hbox{$\buildrel<\over\sim$}}\ }

\begin{document}
\title*{Extragalactic Distances from Planetary Nebulae}
\toctitle{Distances from Planetary Nebulae}
%
%
\titlerunning{Distances from Planetary Nebulae}
%
\author{Robin Ciardullo}
\authorrunning{Robin Ciardullo}
%
%
\institute{The Pennsylvania State University,
           Department of Astronomy and Astrophysics,
           525 Davey Lab
           University Park, PA 16803, USA}

\maketitle              

\begin{abstract}
The [O~III] $\lambda 5007$ planetary nebula luminosity function (PNLF)
occupies an important place on the extragalactic distance ladder.  Since
it is the only method that is applicable to all the large galaxies of 
the Local Supercluster, it is uniquely useful for cross-checking results
and linking the Population~I and Population~II distance scales.  We review
the physics underlying the method, demonstrate its precision, and illustrate
its value by comparing its distances to distances obtained from Cepheids and
the Surface Brightness Fluctuation (SBF) method.  We use the Cepheid and
PNLF distances to 13 galaxies to show that the metallicity dependence of the
PNLF cutoff is in excellent agreement with that predicted from theory, and
that no additional systematic corrections are needed for either method. 
However, when we compare the Cepheid-calibrated PNLF distance scale with the
Cepheid-calibrated SBF distance scale, we find a significant offset: although
the relative distances of both methods are in excellent agreement, the PNLF
method produces results that are systematically shorter by $\sim 15\%$.
We trace this discrepancy back to the calibration galaxies and show how a 
small amount of internal reddening can lead to a very large systematic error.  
Finally, we demonstrate how the PNLF and Cepheid distances to NGC~4258 argue 
for a short distance to the Large Magellanic Cloud, and a Hubble Constant
that is $\sim 8\%$ larger than that derived by the {\sl HST\/} Key Project.

\end{abstract}

\section{Introduction}
The brightest stars have been used as extragalactic distance indicators ever
since the days of Edwin Hubble \cite{hubble}.  However, it was not until the
early 1960's that it was appreciated that young planetary nebulae (PNe) also 
fall into the ``brightest stars'' category.  In their early stages of 
evolution, planetary nebulae are just as luminous as their asymptotic
giant branch (AGB) progenitors; the fact that most of their continuum emission
emerges in the far ultraviolet, instead of the optical or near infrared,
in no way affects their detectability.  On the contrary, because most of
the central star's flux comes out at energies shortward of 13.6~eV, the
physics of photoionization guarantees that this energy is reprocessed 
into a series of optical, IR, and near-UV emission lines.  In fact, $\sim 10\%$ 
of the flux emitted by a young, planetary nebula comes out in a single 
emission line of doubly-ionized oxygen at 5007~\AA.  Thus, for cosmological
purposes, a PN can be thought of as a cosmic apparatus which transforms
continuum emission into monochromatic flux.

\begin{figure}[t]
\begin{center}
\includegraphics[width=1.0\textwidth]{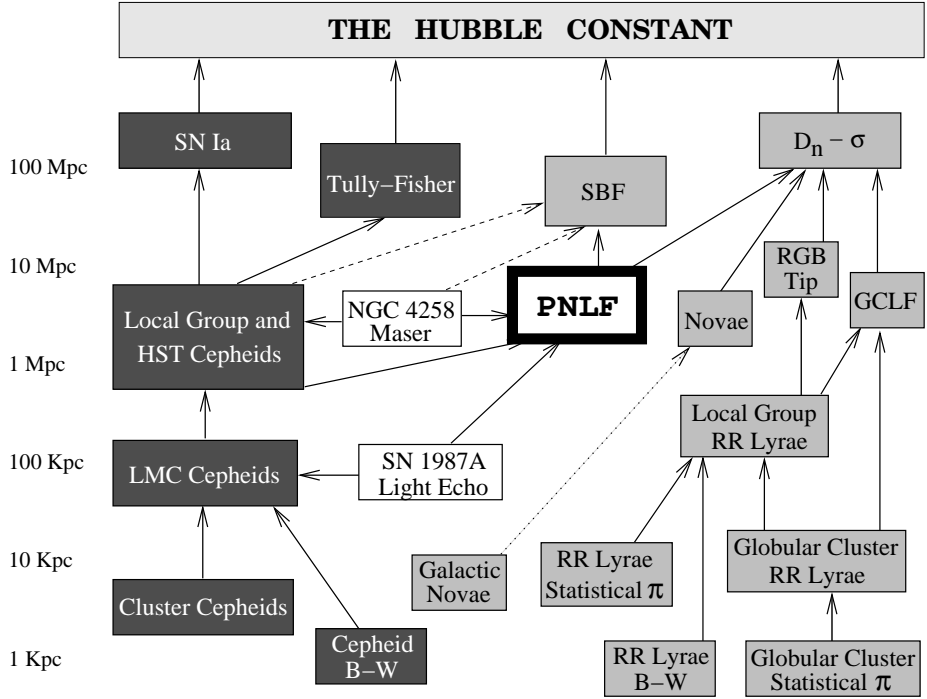}
\end{center}
\caption[]{The extragalactic distance ladder.  The dark boxes show 
techniques useful in star-forming galaxies, the lightly-filled boxes give 
methods that work in Pop~II systems, and the open boxes represent 
geometric distance determinations.  Uncertain calibrations are noted as
dashed lines.  The PNLF is the only method that is equally effective in all
the populations of the Local Supercluster.}
\label{eps1}
\end{figure}

Although the idea of using PNe as standard candles was first presented in
the early 1960's \cite{hw63,hodge}, it was not until the late 1970's that
pioneering efforts in the field were made.  Ford and Jenner \cite{fj78} had
noticed that the visual magnitudes of the brightest planetary nebulae in
M31, M32, NGC~185, and NGC~205 were the same to within $\sim 0.5$~mag.  This
suggested that bright planetary nebulae could be used as standard candles. 
Based on this premise, crude PN-based distances were obtained to M81 
\cite{fj78}, NGC~300 \cite{lg83}, and even several Local Group dwarfs
\cite{jl81}.  These distance estimates were not very persuasive, since at the
time nothing was known about the systematics of bright planetary nebulae or
their luminosity function.  Moreover, it had long been known that Galactic PNe
are definitely {\it not\/} standard candles \cite{berman,minkowski,shklovsky}.
(It is an irony of the subject that in the Milky Way, factor of two distance 
errors are the norm \cite{cks,vdsz,zhang,bensby,phillips}.)  Thus, it was not
until 1989 when the [O~III] $\lambda 5007$ PN luminosity function (PNLF) was 
modeled \cite{p1}, and compared to the observed PNLFs of M31 \cite{p2}, M81 
\cite{p3}, and the Leo~I Group \cite{p4}, that PNe became generally accepted 
as a distance indicator.   Today, the [O~III] $\lambda 5007$ PNLF is one of 
the most important standard candles in extragalactic astronomy, and the only 
method that can be applied to all the large galaxies of the Local Supercluster,
regardless of environment or Hubble type (see Fig.~\ref{eps1}).

\section{Planetary Nebula Identifications}
PNLF observations begin with the selection of a narrow-band filter.  
Ideally, this filter should be centered at 5007~\AA\ at the redshift of the
target galaxy and be 25~\AA\ to 50~\AA\ wide.  Narrower filters may miss
objects that are redshifted out of the filter's bandpass by the galaxy's 
internal velocity dispersion, while broader filters admit too much continuum 
light and invite contamination by [O~III] $\lambda 4959${}.  One subtlety of 
the process is that the characteristics of the filter at the telescope will not
be the same as those in the laboratory.   The central wavelength of an 
interference filter typically shifts $\sim 0.2$~\AA\ to the blue for every 
$1^\circ$~C drop in temperature.  In addition, fast telescope optics will 
lower the filter's peak transmission, shift its central wavelength to the
blue, and drastically broaden its bandpass \cite{eather}.  The observer must 
consider these factors when planning an observation, since without an accurate 
knowledge of the filter transmission curve, precise PN photometry is not 
possible.

\begin{figure}[t]
\begin{center}
\includegraphics[width=1.0\textwidth]{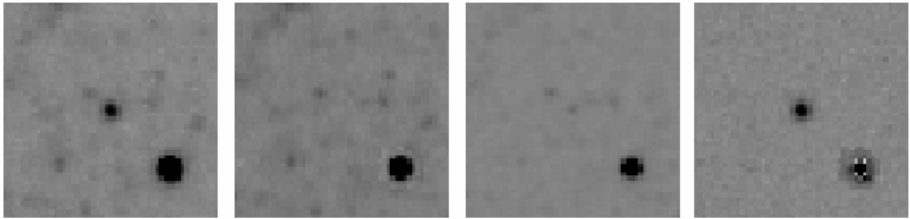}
\end{center}
\caption[]{The first three panels show images of a PN in NGC~2403 in [O~III] 
$\lambda 5007$, continuum $\lambda 5300$, and H$\alpha$.  The last column
displays the [O~III] on-band minus off-band difference image.  The PN 
candidate is in the middle of the frame.  All PNe in the top $\sim 1$~mag
of the PNLF are stellar, invisible in the continuum, and much brighter in 
[O~III] $\lambda 5007$ than H$\alpha$.}
\label{eps2}
\end{figure}

PN observations in early-type galaxies are extremely simple.  One images the
galaxy through the narrow on-band filter, and then takes a similar image
through a broader, off-band filter.  The two frames are then compared,
either by ``blinking'' the on-band image against the off-band image, or by
creating an on-band minus off-band ``difference'' frame.  Point sources 
which appear on the on-band frame, but are completely invisible on the 
offband frame, are planetary nebula candidates (see Fig.~\ref{eps2}).  In 
this era of wide-field mosaic CCD cameras, $V$ filters are often used in place 
of true off-band filters.  This works for most extragalactic programs, but is 
not ideal.  Since the $V$-band includes the 5007~\AA\ emission line, its use
as an  ``off-band'' may cause bright PNe to appear (faintly) in the continuum.
Photometric techniques which use the difference image will therefore be 
compromised.

\begin{figure}[t]
\begin{center}
\includegraphics[width=1.0\textwidth]{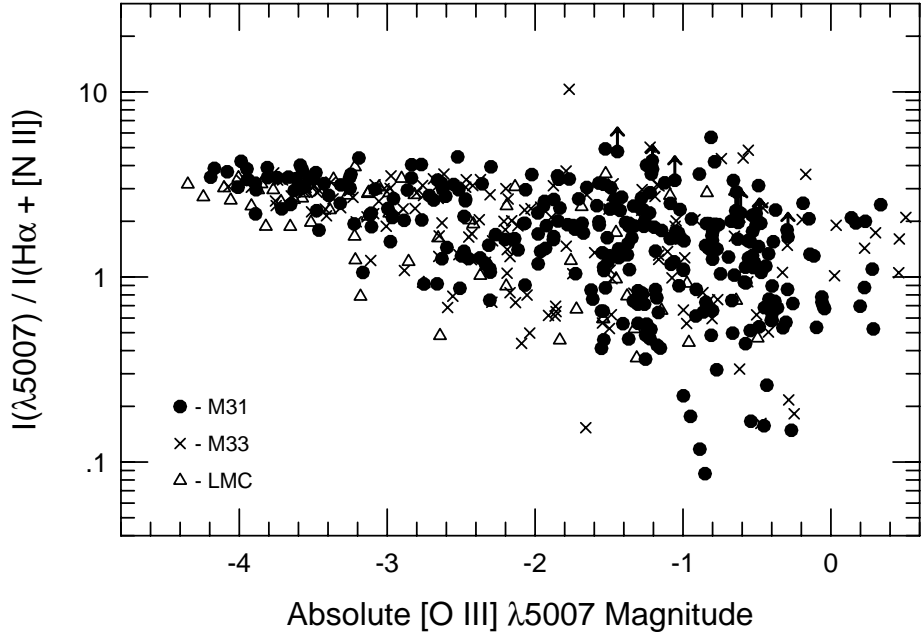}
\end{center}
\caption[]{The [O~III] $\lambda 5007$ to H$\alpha$+[N~II] line ratio for PNe 
in the bulge of M31 \cite{p2}, the disk of M33 \cite{dur02}, and the Large 
Magellanic Cloud \cite{md1,md2,vdm,p6}.  This line ratio is useful for 
discriminating bright PNe from compact H~II regions.}
\label{eps3}
\end{figure}

Since virtually every [O~III] $\lambda 5007$ source in an elliptical or
lenticular galaxy is a planetary nebula, PNLF measurements in these
systems are straightforward.  However, in spiral and irregular galaxies, 
this is not the case.  H~II regions and supernova remnants are also strong 
[O~III] $\lambda 5007$ emitters, and in late-type systems, these objects can 
numerically overwhelm the planetaries.  Fortunately, most H~II regions are 
resolvable (at least in galaxies closer than $\sim 10$~Mpc), whereas 
extragalactic PNe, which are always less than 1~pc in radius \cite{acker}, are 
stellar.  Thus, any object that is not a point source can immediately be
eliminated from the sample.  To remove the remaining contaminants, one can use 
H$\alpha$ as a discriminant.  Planetary nebulae inhabit a distinctive region of
[O~III] $\lambda 5007$-H$\alpha$ emission-line space.  As illustrated in 
Fig.~\ref{eps3}, objects in the top magnitude of the PNLF all have $\lambda 
5007$ to H$\alpha$+[N~II] line ratios greater than $\sim 2$.  This is in 
contrast to H~II regions, which typically have ratios less than one
\cite{shaver}.  This difference in excitation is an effective diagnostic for 
removing whatever compact H~II regions remain in the sample.

There are two other sources of contamination which may occur in deep planetary
nebula surveys.  The first is background galaxies.  At
$z = 3.12$, Ly$\alpha$ is redshifted in the bandpass of the [O~III] $\lambda
5007$ filter, and at fluxes below $\sim 10^{-16}$~ergs~cm$^{-2}$~s$^{-1}$,
unresolved and marginally resolved galaxies with extremely strong Ly$\alpha$
emission (equivalent widths $\gtrsim 300$~\AA\ in the observers frame) do
exist \cite{hu,kudritzki,freeman}.  Fortunately, the density of these 
extraordinary objects is relatively low, $\sim 1$~arcmin$^{-2}$~per unit 
redshift interval brighter than $5 \times 10^{-17}$~ergs~cm$^{-2}$~s$^{-1}$
\cite{blank}.  Thus while an occasional high-redshift interloper may 
be found within the body of a galaxy \cite{m51}, these objects are unlikely to 
distort the shape of the luminosity function.

The second source of confusion is specific to the Virgo Cluster.  Between 10\%
and 20\% of the stellar mass of rich clusters lies outside of any galaxy
in intergalactic space \cite{ipn1,irgb,kara}.  PN surveys within these 
systems will therefore be contaminated by intracluster objects.  In clusters 
such as Fornax, where the line-of-sight thickness is small \cite{p9,tonry},
the effect of intracluster planetaries on the target galaxy's PNLF is minimal.  
However, the Virgo Cluster's depth is substantial \cite{p5,yfo,wb,magda}, so
surveys in this direction will contain a significant number of foreground
sources.  These intracluster objects can distort the galactic PNLF and 
possibly produce a biased distance estimate.  The best way to minimize
the effect is to limit PN surveys to the inner regions of galaxies
(where the ratio of galactic to intracluster light is high), or statistically
subtract the contribution of intracluster objects \cite{m87pn}.

\section{Deriving Distances}
Once the PNe are found, the next step is to measure their brightnesses and
define a statistically complete sample.  The first step is easy.  A 
significant advantage of the PNLF method is that it does not require complex
crowded-field photometric algorithms.  Raw instrumental magnitudes can be
derived using simple aperture photometry or point-spread-function fitting 
procedures, and then turned into monochromatic [O~III] $\lambda 5007$ fluxes 
using the techniques described in \cite{jqa}.  These fluxes are usually quoted 
in terms of magnitudes via
\begin{equation}
m_{5007} = -2.5 \log F_{5007} - 13.74
\end{equation}
The zero point of this system is not completely arbitrary.  In this 
``standard'' system, a PN's $\lambda 5007$ magnitude is roughly equal to the
magnitude it would have if viewed through the broadband $V$ filter \cite{p1}. 
Bright PNe in M31 have $m_{5007} \sim 20$, while the brightest planetaries in 
Virgo have $m_{5007} \sim 26.5$.

The determination of statistically complete samples can be more time consuming.
Although the onset of incompleteness can be found via the ``traditional''
method of adding artificial stars to frames and measuring the recovery 
fraction, there is a short cut.   Experiments have shown that PN counts are
not affected by incompleteness until the recorded signal-to-noise drops
below a threshold value of $\sim 10$ \cite{cfnjs,hui93}.  Since extragalactic
PNe are faint, this means that the probability of PN detection is a 
function of two parameters:  the instrumental magnitude of the planetary,
and the brightness of the underlying background.  In early-type systems,
where the galactic background is smooth and well-behaved, the creation of 
a statistical sample is therefore straightforward.  One chooses an
isophote and uses the signal-to-noise threshold to calculate the completeness
limit (see \cite{p3,p4}).  In spiral and irregular galaxies, where the 
underlying background is irregular and complex, the process is more
empirical: one selects the brightest (most uncertain) background in the sample, 
and uses the signal-to-noise each PN would have if it were projected on
that background \cite{p11}.  In either case, the limiting magnitude for 
completeness need not be precise.  The PNLF method depends far more on the 
brightest objects in the sample than the dimmest; small errors at the 
faint end of the luminosity function have little effect on the final derived
distance.  

Once a statistical sample of planetaries has been defined, PNLF distances are 
obtained by fitting the observed luminosity function to an empirical law.
For simplicity, Ciardullo \etal \cite{p2} have fit the bright-end cutoff
with the function
\begin{equation}
N(M) \propto e^{0.307 M} \{ 1 - e^{3 (M^* - M)} \}
\end{equation}
though other forms of the relation are possible \cite{men93}.  In the above
equation, the key parameter is $M^*$, the absolute magnitude of the brightest
possible planetary nebula.  Despite some efforts at Galactic calibrations
\cite{pot90,men93}, the PNLF remains a secondary standard candle.  The 
original value for the zero point, $M^* = -4.48$, was based on an M31 
infrared Cepheid distance of 710~kpc \cite{welch} and a foreground extinction 
of $E(B-V)=0.11$ \cite{mcclure}.  Since then, M31's distance has increased 
\cite{keyfinal}, its reddening has decreased \cite{schlegel}, and, most 
importantly, the Cepheid distances to 12 additional galaxies have been 
included in the calibration \cite{p12}.  Somewhat fortuitously, the current 
value of $M^*$ is only 0.01~mag fainter than the original value, $M^* = -4.47$ 
\cite{p12}.

Before proceeding further, it is important to note that equation (2) only
seeks to model the top $\sim 1$~mag of the PN luminosity function.  At
fainter magnitudes, large population-dependent differences exist.  For example,
in M31's bulge the PNLF monotonically increases according to the exponent in
the empirical law \cite{p2,hui94}.  However the luminosity functions of
the Small Magellanic Cloud and M33 are not so well-behaved: compared to M31, 
these galaxies are a factor of $\sim 2$ deficient in PNe in the magnitude range 
$-2 < M_{5007} < +2$ \cite{jd02,dur02}. Fortunately, this behavior (which
depends on the system's star-formation history and is easily explained in terms
of stellar evolution and photoionization theory) does not affect the bright end
of the PNLF{}. It is therefore irrelevant for PNLF distance determinations.

Finally, before any distance can be derived, one must consider the effect
of extinction on the distance indicator.  For PNLF observations, the ratio of
total to differential extinction is non-negligible ($A_{5007} = 3.5 E(B-V)$ 
\cite{cardelli}), so this issue has some importance.   There are two 
sources to consider: foreground extinction from the Milky Way, and internal 
extinction from the program galaxy.  The former quantity is readily obtainable 
from reddening maps derived from H~I measurements and galaxy counts \cite{bh84}
and/or from the DIRBE and IRAS satellite experiments \cite{schlegel}.
However, the latter contribution to the total extinction is more
problematic.  In the Galaxy, the scale height of PNe is significantly larger
than that of the dust \cite{mb81}.  If the same is true in other galaxies,
then we would expect the bright end of the PNLF to always be dominated 
by objects in front of the dust layer.  This conclusion seems to be supported
by observational data \cite{p11,p12} and numerical models \cite{p11}, both
of which suggest that the internal extinction which affects a galaxy's PN 
population is $\lesssim 0.05$~mag.  We will, however, revisit this issue in 
Section~5.

\section{Why the PNLF Works}

The effectiveness of the PNLF technique has surprised many people.  After all,
a PN's [O~III] $\lambda 5007$ flux is directly proportional to the luminosity
of its central star, and this luminosity, in turn, is extremely sensitive to 
the central star's mass.  Since the distribution of PN central star masses 
depends on stellar population via the initial mass-final mass relation
\cite{weidemann}, one would think that the PNLF cutoff would be 
population dependent.

Fortunately, this does not appear to be the case, and, in retrospect, the
invariance is not difficult to explain.   First, consider the question of
metallicity.  The [O~III] $\lambda 5007$ flux of a bright planetary is
proportional to its oxygen abundance, but since $\gtrsim 10\%$ of the central
star's flux comes out in this one line, the ion is also the nebula's primary 
coolant.  Consequently, if the abundance of oxygen is decreased, the nebula's 
electron temperature will increase, the number of collisional excitations per 
ion will increase, and the amount of emission per ion will increase.  The 
result is that the flux in [O~III] $\lambda 5007$ depends only on the square 
root of the nebula's oxygen abundance \cite{p1}.

Meanwhile, the PN's core reacts to metallicity in the opposite manner.
According to models of AGB and post-AGB evolution \cite{lattanzio,brocato}
if the metal abundance of a star is decreased, then the bound-free opacity
within the star will decrease, and the emergent UV flux will increase.  This
will cause additional energy to be deposited into the nebula, and increase
the amount of [O~III] $\lambda 5007$ emission.  Since this effect is small, and 
works in the opposite direction as the nebular dependence, the overall result
is that the bright-end cutoff of the PNLF should be almost independent of 
metallicity.

A more sophisticated analysis by Dopita, Jacoby, \& Vassiliadis \cite{djv}
confirms this behavior.  According to their models, the dependence of
$M^*$ on metallicity is weak and non-monotonic; a quadratic fit to the
relation yields
\begin{equation}
\Delta M^* = 0.928 {\rm [O/H]}^2 + 0.225 {\rm [O/H]} + 0.014
\end{equation}
where [O/H] is the system's logarithmic oxygen abundance referenced to the
solar value of $12+\log {\rm (O/H)} = 8.87$ \cite{grevesse}.  Inspection
of equation (3) reveals that $M^*$ is brightest when the population's
metallicity is near solar.  In super-metal rich systems $M^*$ fades, but
since all metal-rich galaxies contain substantial populations of metal-poor
stars, this part of the metallicity dependence should not be observed.
Moreover, although $M^*$ also fades in metal-poor systems, the change is 
gradual, so as long as the oxygen abundance of the host galaxy is $12 + 
\log {\rm (O/H)}~\gtrsim 8.3$ (\ie greater than two-thirds that of the LMC),
the effect on distance determinations should be less than 10\%.  This weak 
dependence on metallicity is one reason why PNLF distances are so robust.

\begin{figure}[t]
\begin{center}
\includegraphics[width=1.0\textwidth]{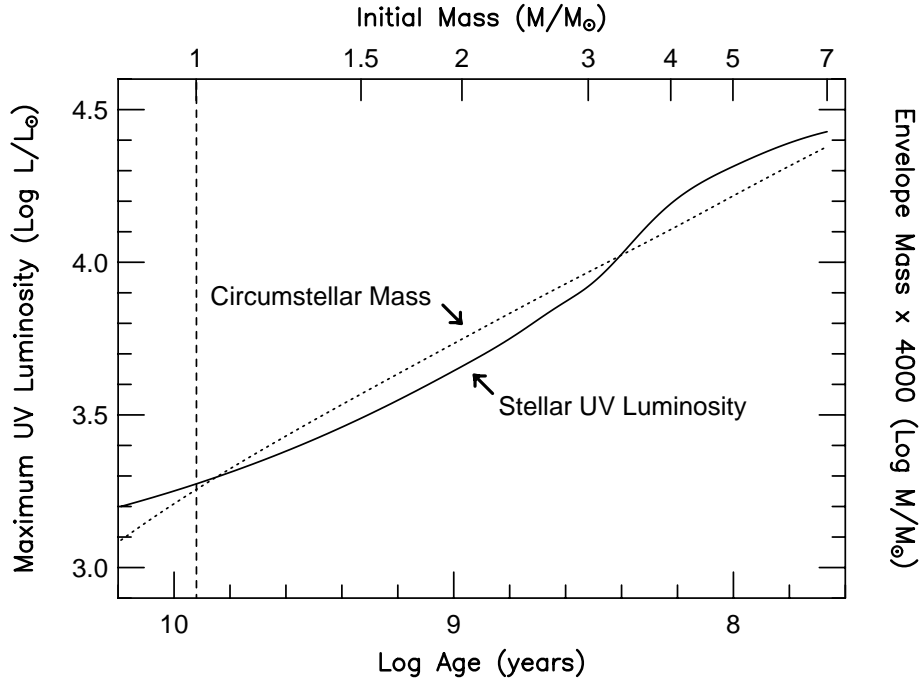}
\end{center}
\caption[]{A comparison of the maximum amount of ionizing radiation emitted
by PN central stars against the mass of the stars' envelopes.  The curves
assume that the central stars are hydrogen burners \cite{blocker} and
use the Wiedemann initial-mass final-mass relation \cite{weidemann} with 
minimal RGB mass loss.   The approximate lower-mass limit for PN progenitors 
is noted by a dotten line \cite{j97}; the conversion between initial mass
and age comes from Iben and Laughlin \cite{iben}.  The similarity of the
relations implies that extinction will act to suppress the [O~III] $\lambda 
5007$ emission from high core-mass planetaries.}
\label{eps4}
\end{figure}

The reaction of the PNLF cutoff to population age is slightly less obvious,
but no more complicated.  Post-AGB evolutionary models \cite{vw94,blocker}
predict that the maximum luminosity and temperature achieved by a PN's
central star is highly dependent on its core mass, with (very roughly)
$L \propto M^3$ and $T_{\rm max} \propto M^{2.5}$ for intermediate-mass
hydrogen burning models.  Consequently, high-mass central stars should be
extremely bright in the UV and their nebulae should be exceptionally 
luminous in [O~III] $\lambda 5007$.  Since the mass of a central star is
proportional to the mass of its progenitor (through the initial-mass 
final-mass relation \cite{weidemann}), this line of reasoning seems to imply
the existence of some extremely luminous Population~I planetaries. 
In fact, these over-luminous objects do exist.  In the Magellanic Clouds, 9 out 
of the 74 planetaries with well-calibrated spectrophotometry \cite{md1,md2,p6} 
have intrinsic [O~III] $\lambda 5007$ magnitudes brighter than $M^*$. 
Conversely, in the central regions of M31, where the bulge population 
dominates, only one out of 12 spectrophotometrically observed PNe is 
superluminous in [O~III] \cite{jc99}.  However, {\it in every case,} these 
over-luminous objects are heavily extincted by circumstellar material, so 
that no PN has an observed [O~III] $\lambda 5007$ flux brighter than $M^*$.

\begin{figure}[t]
\begin{center}
\includegraphics[width=1.0\textwidth]{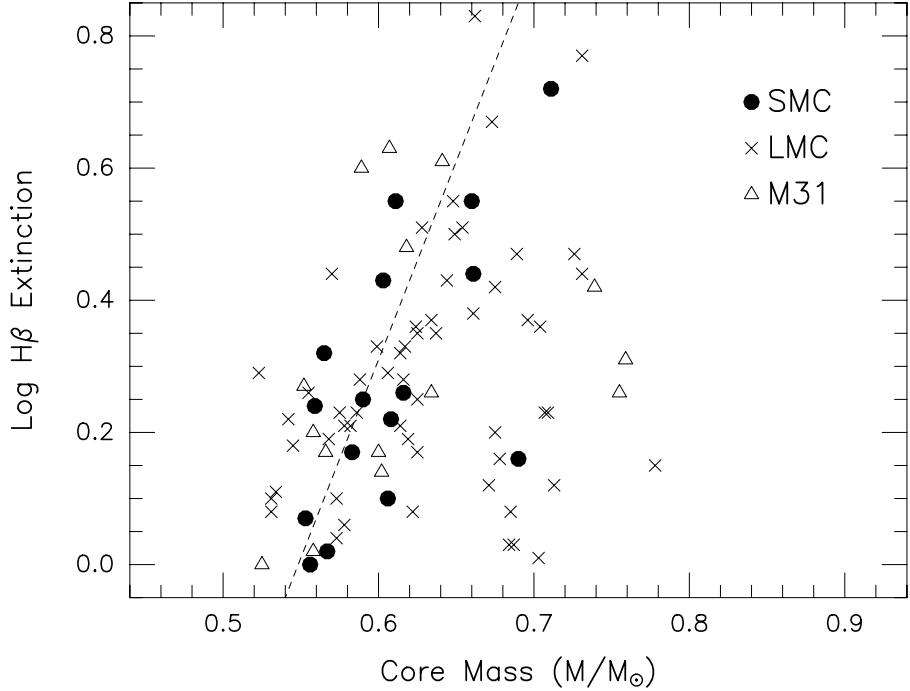}
\end{center}
\caption[]{The correlation between circumstellar extinction and central star
mass for planetary nebulae in the Magellanic Clouds and M31.  The 
extinction values are based on the Balmer decrement; the core masses have
been derived via comparisons with hydrogen-burning evolutionary tracks.  
The slope of the relation is $5.7 \pm 0.7$ for the SMC, $6.3 \pm 1.3$ for 
the LMC, and $8.5 \pm 1.6$ for M31.}
\label{eps5}
\end{figure}

In order to understand this phenomenon, one needs to consider the ratio of
a nebula's input energy to its own circumstellar extinction.  The former
quantity is dictated by the central star's flux shortward of 912~\AA, which
via the initial-mass final-mass relation, depends sensitively on the mass of
the star's progenitor.  The latter value is proportional to the amount of mass
lost during the star's AGB phase, which is also set by the progenitor mass.
Figure~\ref{eps4} compares these two values at the time when the central star's
UV flux is greatest.   Remarkably, the two functions are extremely similar
throughout the entire range of progenitor masses.  If the efficiency of 
circumstellar extinction is the same for all planetaries, then the figure
implies that $M^*$ will be independent of population age to within 
$\sim 0.2$~mag.  Since self-extinction is probably more efficient around
high-mass cores (since their faster evolutionary timescales give the 
material less time to disperse), this simple analysis suggests that
high-mass PNe which are intrinsically more luminous than $M^*$ will always be 
extincted below the empirical PNLF cutoff.  

Observational support for this scenario is shown in Fig.~\ref{eps5}, which 
plots the relation between PN core mass and circumstellar extinction for 
[O~III]-bright planetaries in the LMC, the SMC, and M31 \cite{cj99}.  The core 
masses of Fig.~\ref{eps5} have been derived by placing the central stars on the 
HR diagram (via photoionization modeling of the PNe's emission lines), and 
comparing their positions to the evolutionary tracks of hydrogen-burning
post-AGB stars \cite{blocker}; the plotted extinction estimates have
been inferred from the PNe's Balmer line ratios.  Since the derived 
temperatures and luminosities of central stars have some uncertainty, and a 
fraction of PNe will be burning helium instead hydrogen, a good amount of 
scatter in the diagram is expected.  Nevertheless, there is a statistically 
significant correlation between core mass and circumstellar extinction for the 
PN populations of all three galaxies.  The best-fitting slope of 
$\sim 6$~mag/$M_{\odot}$ more than compensates for the increased UV luminosity 
associated with the high-mass cores.  In fact, when combined with the 
initial-mass final-mass relation \cite{weidemann}, the steep slope of 
Figure~\ref{eps5} predicts that $M^*$ should vary by less than $\sim 0.1$~mag 
in all populations older than 0.4~Gyr \cite{cj99}.  In younger populations, 
$M^*$ may fade, but since all galaxies contain at least some stars older than 
$\sim 0.4$~Gyr, this behavior should not be observable.  The value of 
$M^*$ in a star-forming galaxy should therefore be the same as that of
an old stellar population.

\section{Tests of the Technique}
In the past decade, the PNLF has been subjected to a number of rigorous
tests.  In general, these tests fall into four categories.

\subsection{Internal Tests Within Galaxies}
The first and perhaps simplest test applied to the PNLF involves taking 
advantage of population differences within galaxies.  Spiral galaxies have 
significant metallicity gradients \cite{zkh}, and the stellar population of a 
spiral's bulge is certainly different from that of its disk and halo. 
Population differences exist in elliptical galaxies as well, as their radial 
color gradients attest \cite{peletier}.  If one can measure the distance to a 
sample of planetaries projected close to a galaxy's nucleus, and then do the 
same for PN samples projected at intermediate and large galactocentric radii, 
one can determine just how sensitive the PNLF is to changes in stellar 
population.  

\begin{figure}
\begin{center}
\includegraphics[width=1.0\textwidth]{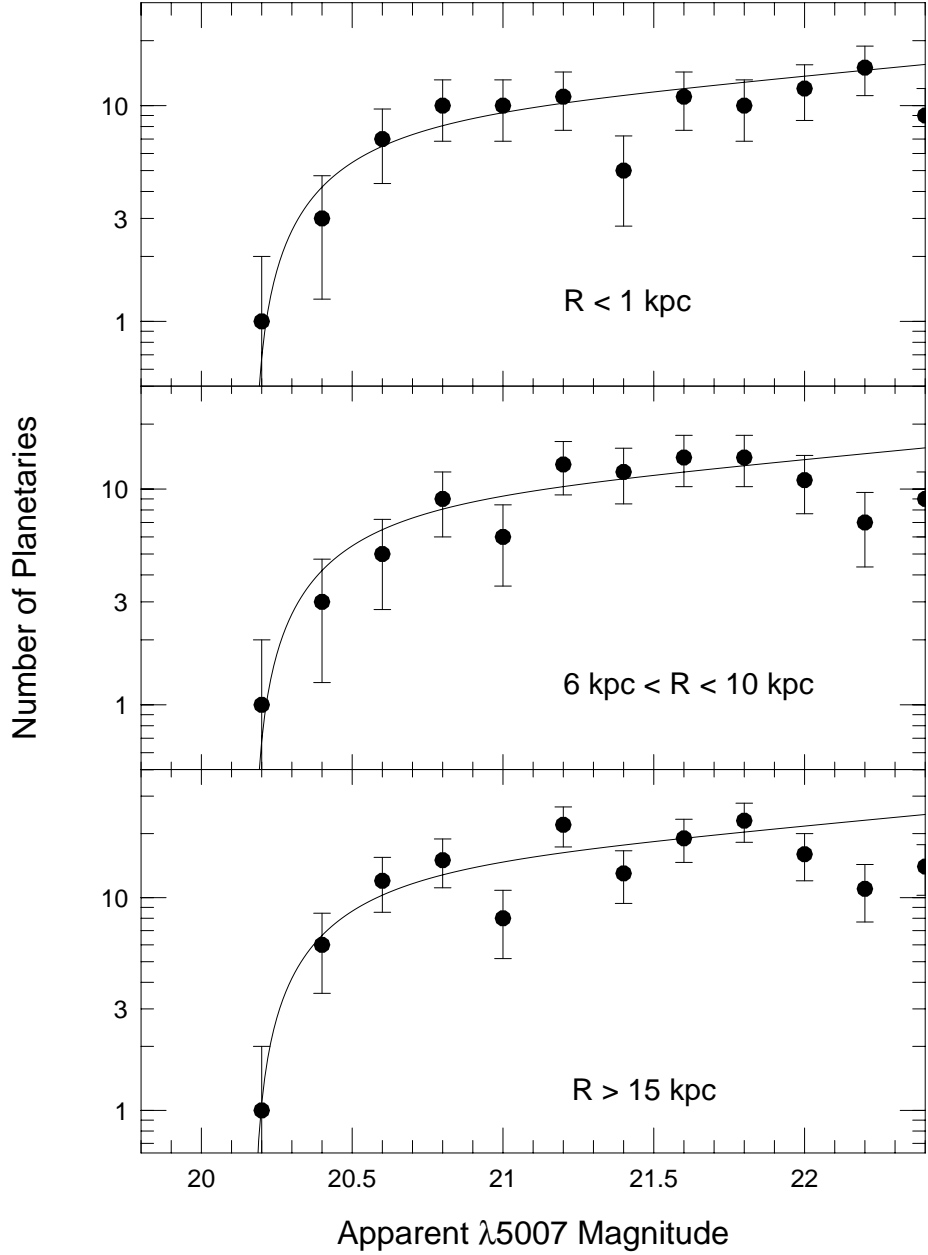}
\end{center}
\caption[]{The observed planetary nebula luminosity functions for samples of
M31 PNe projected at three different galactocentric radii.  The curves show 
the best-fitting empirical law.  The derived PNLF distances are consisted
to within $\sim 0.05$~mag.  The turnover in the luminosity function past
$m_{5007} \, \gtrsim 22$ in the intermediate and large-radii samples is 
real, and indicates the presence of relatively massive PN central stars.}
\label{eps6}
\end{figure}

Four galaxies now have large enough PN samples for this test:  two Sb spirals
(M31 \cite{hui94} and M81 \cite{mag81}), one large elliptical (NGC~4494
\cite{p10}), and one blue, interacting elliptical (NGC~5128 \cite{hui93}).  
The data for M31 are shown in Fig.~\ref{eps6}.  No significant change in the 
PNLF cutoff has been observed in any of these objects.  Given the diversity of 
stellar populations sampled, this result, in itself, is impressive proof of
the robustness of the method.

\subsection{Internal Tests within Clusters}
A second internal test of the PNLF uses multiple galaxies within a common 
cluster.  Galaxy groups are typically $\sim 1$~Mpc in diameter.  PNLF distances
to individual cluster members should therefore be consistent to within this 
value.  Moreover, if the technique really is free of systematic errors, the 
measured distances should be uncorrelated with any galactic property, such as 
color, luminosity, metallicity, or Hubble type.  

\begin{figure}[t]
\begin{center}
\includegraphics[width=1.0\textwidth]{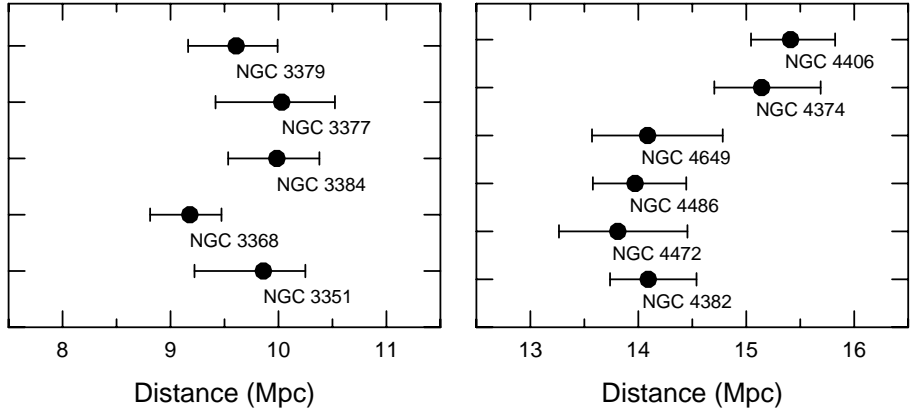}
\end{center}
\caption[]{PNLF distance measurements to the Leo~I Group (left) and the
Virgo Cluster (right).  The Leo~I galaxies possess a range of Hubble
types from SBb to E0; the Virgo galaxies are all ellipticals or lenticulars,
but range in color from $1.28 < (U-V) < 1.64$.  The PNLF measurements in Leo~I
place all the galaxies within $\sim 1$~Mpc of each other, while in Virgo,
the method easily resolves the background galaxies NGC~4374 and 4406 from the
main body of the cluster.}
\label{eps7}
\end{figure}

To date six galaxy clusters have multiple PNLF measurements:  the M81 Group
(M81 and NGC~2403 \cite{p3,p11}), the NGC~1023 Group (NGC~891 and 1023 
\cite{p7}), the NGC~5128 Group (NGC~5102, 5128, and 5253 
\cite{mcmillan,hui93,n5253}), the Fornax Cluster (NGC~1316, 1399, and 1404 
\cite{p9}), the Leo~I Group (NGC~3351, 3368, 3377, 3379, and 3384 
\cite{p12,p11,p4}), and the Virgo Cluster (NGC~4374, 4382, 4406, 4472, 4486,
and 4649 \cite{p5}).  In each system, the observed
galaxies have a range of color, absolute magnitude, and Hubble type.  In
none of the clusters is there any hint of a systematic trend.  Indeed, as
Fig.~\ref{eps7} indicates,  PNLF measurements in Virgo easily resolve the
M84/M86 Group, which is falling into the main body of Virgo from 
behind \cite{bohringer}.

\subsection{Comparisons with Cepheid Distances}

Perhaps the most interesting test one can perform for any distance indicator
is to compare its results to those of other methods.  Such tests are crucial
to the scientific method.  While consistency checks, such as those described 
above, provide important information on the systematic behavior of a standard 
candle, external comparisons are the only way to assess the total uncertainty 
associated with a given rung of the distance ladder.  

\begin{figure}[t]
\begin{center}
\includegraphics[width=1.0\textwidth]{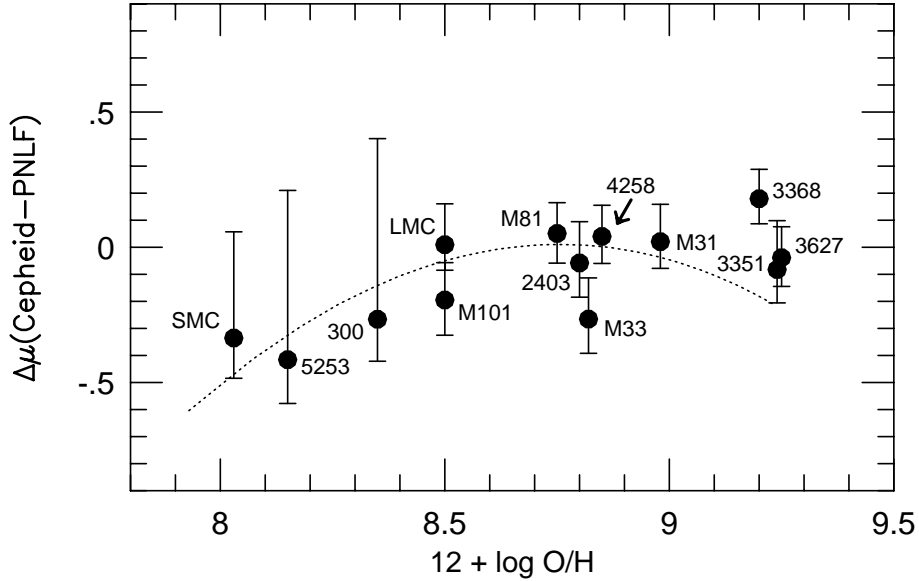}
\end{center}
\caption[]{A comparison of the PNLF and Cepheid distance moduli as function
of galactic oxygen abundance, as estimated from the systems' H~II regions 
\cite{fdatabase}.  No metallicity correction has been applied to either
distance indicator.  The error bars represent the formal uncertainties of
the methods added in quadrature; small galaxies with few PNe have generally
larger errors.  The curve shows the expected reaction of the PNLF to
metallicity \cite{djv}.  Note that metal-rich galaxies should not follow 
this relation, since these objects always contain enough low metallicity 
stars to populate the PNLF's bright-end cutoff.  The agreement between the
two distance estimators is excellent, and the scatter is consistent with the 
internal errors of the methods.}
\label{eps8}
\end{figure}

Figure~\ref{eps8} compares the PNLF distances of 13 galaxies (derived
using the foreground extinction estimates from DIRBE/IRAS \cite{schlegel})
with the final Cepheid distances produced by the {\sl HST Key Project\/}
\cite{keyfinal}.  Neither set of numbers has been corrected for the
effects of metallicity.   Since the absolute magnitude of the PNLF cutoff,
$M^*$, is based on these Cepheid distances, the weighted mean of the 
distribution must, by definition, be zero.  However, the residuals about this 
mean, and the systematic trends in the data, are valid indicators of the 
accuracy of the measurements.

As Fig.~\ref{eps8} illustrates, the scatter between the Cepheid distances
and the PNLF distances is impressively small.  Except for the most 
metal-poor systems, the residuals are perfectly consistent with the internal 
uncertainties of the methods.  Moreover, the systematic shift seen at 
low-metallicity is exactly that predicted by PNLF theory \cite{djv}.   
If $M^*$ were to be corrected using equation (3), the systematic error would 
completely disappear.  This excellent agreement strongly suggests that neither 
the PNLF nor the Cepheid measurements need further metallicity corrections.

\subsection{Comparisons with Surface Brightness Fluctuations}
Another instructive comparison involves distances derived from the measurement
of Surface Brightness Fluctuations (SBF) \cite{tonry}.  SBF distances have a
precision comparable to that of the PNLF, but the technique can only be applied
to smooth stellar populations, such as those found in elliptical and lenticular
galaxies.  Like the PNLF, SBF distances rely on Cepheid measurements for 
their calibration; consequently, a comparison of the two indicators
gives a true measure of the external error associated with climbing a rung
of the distance ladder.

\begin{figure}[t]
\begin{center}
\includegraphics[width=1.0\textwidth]{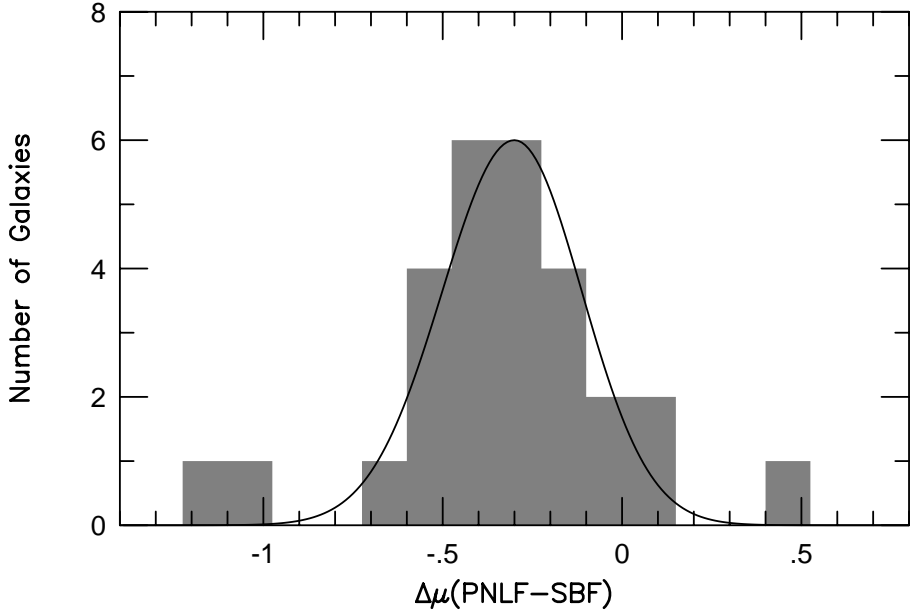}
\end{center}
\caption[]{A histogram of the difference between the PNLF and SBF
distance moduli for 28 galaxies measured by both methods.  The two worst
outliers are the edge-on galaxies NGC~4565 ($\Delta\mu = -0.80$) and
NGC~891 ($\Delta\mu = +0.71$).  NGC~4278 is also an outlier ($\Delta\mu =
-0.70$).  The curve represents the expected dispersion of the data.
The figure demonstrates that the absolute scales of the two techniques are
discrepant, but the internal and external errors of the methods agree.}
\label{eps9}
\end{figure}

To date, 28 galaxies have been measured with both the PNLF and SBF methods.  A
histogram of the distance residuals is shown in Figure~\ref{eps9}.  There are
three important features to note.

The first interesting property displayed in the figure is the presence 
of three obvious outliers.  The two worst offenders are NGC~4565 ($-0.8$~mag 
from the mean) and NGC~891 ($+0.7$~mag from the mean).  Both are edge-on 
spirals --- the only two edge-on spirals in the sample.  Clearly one (or both)
methods have trouble measuring the distances to such objects.  Given the 
sensitivity of SBF measurements to color gradients, it is likely that the 
problem with these galaxies lies there, but an error in the PNLF technique 
cannot be ruled out.

The second important feature of Fig.~\ref{eps9} involves the scatter 
between the PNLF and SBF distance estimates.  The curve plotted in 
the figure is not a fit to the data: it is instead the {\it expected\/} 
scatter in the measurements, as determined by propagating the uncertainties 
associated with the PNLF distances, the SBF distances, and Galactic reddening. 
It is obvious that the derived curve is in excellent agreement with the data.
This proves that the quoted uncertainties in the methods are reasonable.
It also leaves little room for additional random errors associated with 
measurements.  

\begin{figure}[t]
\begin{center}
\includegraphics[width=1.0\textwidth]{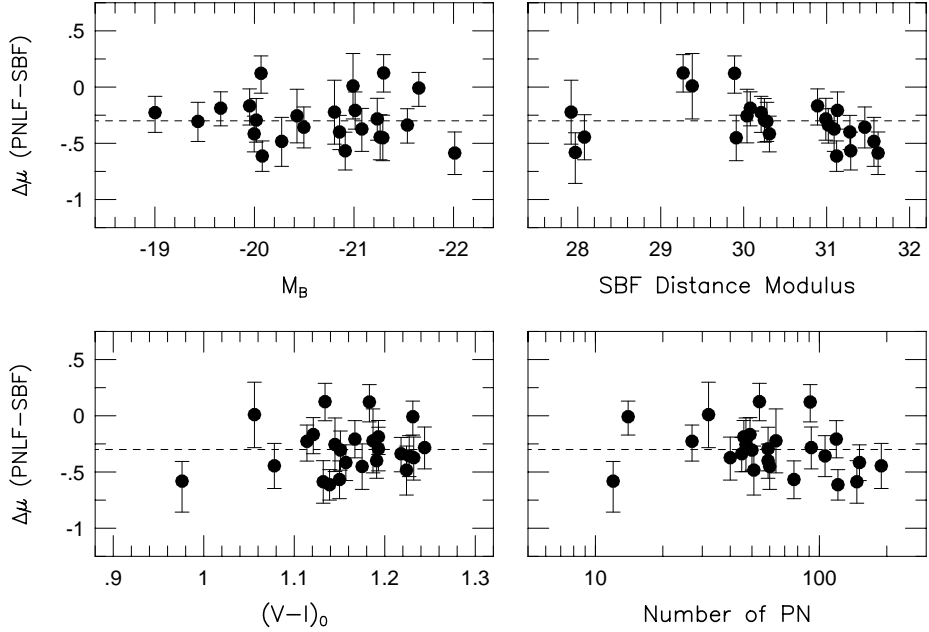}
\end{center}
\caption[]{The difference between SBF and PNLF distance moduli plotted 
against galactic absolute magnitude, distance, color, and number of PNe in
the statistical sample.  The three discrepant galaxies, NGC~891, 4565, and
4278, have not been plotted.  The correlation with SBF distance modulus
is marginally significant ($P \sim 0.1$), due to the low values of the five
most distant objects; if these galaxies are removed from the sample, the
significance of the correlation disappears.  No other correlations exist in
any of the panels.}
\label{eps10}
\end{figure}

The latter conclusion is confirmed in Fig.~\ref{eps10}.  If either method were
significantly affected by population age or metallicity, or if the PNLF
fitting-technique were incorrect, then the PNLF-SBF distance residuals would
correlate with galactic absolute magnitude, color, or PN population.  No
such trend exists.  In fact, the only possible correlation present in the 
figure is with distance: if one {\it only\/} considers galaxies with 
$(m-M)_{\rm SBF} > 30.6$, then the residuals do correlate with distance at the
95\% confidence level.  Such behavior might be expected if the PN samples found 
in distant galaxies were contaminated by background emission-line galaxies (or 
in the case of rich clusters, foreground intracluster stars).  However, if the 
five most distant objects are deleted from the sample, the correlation goes 
away, proving that, in terms of relative distances, the PNLF and SBF techniques
are in excellent agreement.  

Interestingly, the same cannot be said for the methods' absolute distances.
The PNLF zero point comes from planetary nebula observations in the 13 Cepheid 
galaxies displayed in Fig.~\ref{eps8}; the formal uncertainty in $M^*$ is
$\sim 0.05$~mag.  Similarly, the SBF zero point is based on fluctuation
measurements in the bulges of six Cepheid spirals; its estimated uncertainty
is $\sim 0.04$~mag.  If both calibrations were accurate, then the mean
of the PNLF-SBF distance residuals would be $0.0 \pm 0.07$.  It is not: 
as Figs.~\ref{eps9} and \ref{eps10} indicate, SBF distances are, on average 
$0.30 \pm 0.05$~mag larger than PNLF distances.  Even if the five most distant
galaxies are excluded, the remaining $\sim 0.26$~mag offset is more than 
$3 \, \sigma$ larger than expected.  Clearly, there is an important source of 
error that is not being considered by one (or both) techniques. 

The most likely explanation for the discrepancy involves internal extinction
in the Cepheid calibration galaxies.   To calibrate an extragalactic 
standard candle with Cepheids, one needs to measure the apparent brightness
of the candle, $m$, and assume some value for the intervening extinction.
Hence
\begin{equation}
M = m - \mu_{Cep} - R_{\lambda} \, E(B-V)
\end{equation}
where $M$ is the derived absolute magnitude of the object and $R_{\lambda}$ is 
the ratio of total to differential reddening at the wavelength of interest.  
For most methods (including the PNLF), if the reddening to a galaxy is 
underestimated, then the brightness of the standard candle is underestimated,
and the distance scale implied by the observations is underestimated.  
However, in the case of the $I$-band SBF technique, the standard candle,
$\bar M_I$ has a strong color dependence, with $\bar M_I = C + 4.5 (V-I)_0$
\cite{tonry}.  Consequently, the zero-point of the system, $C$, is defined
through
\begin{equation}
C = \bar m_I - \mu_{Cep} - 4.5 (V-I)_{\rm obs} + (4.5 \, R_V - 5.5 \,R_I) \,
E(B-V)
\end{equation}
Because $R_V > R_I$, an underestimate of reddening results in an overestimate
of the brightness of the standard candle, and a distance scale that is too
large.  Since the PNLF and SBF methods react in opposite directions to 
reddening, even a small amount of internal extinction in the bulges of the 
calibrating spirals can lead to a large discrepancy between the systems in the 
exact sense that is seen.  Specifically, if only the SBF measurements are 
affected, then the technique's distance scale will be too large by $4.2 \, 
\sigma_{E(B-V)}$ \cite{cardelli}.   Moreover, if both techniques are affected, 
then $\sigma_{\Delta \mu} = 7.7 \, \sigma_{E(B-V)}$.  With such a large 
coefficient, it would take only a small amount of internal reddening, $E(B-V) 
\sim 0.04$~mag to explain the discrepancy seen in the figures.

If internal extinction really is responsible for the offset displayed in
Fig.~\ref{eps9}, then the zero points of both systems must be adjusted.  These 
corrections will propagate all the way up the distance ladder.  For example, 
according to the {\sl HST Key Project,} the SBF-based Hubble Constant is $69 
\pm 4$ (random) $\pm 6$ (systematic) km~s$^{-1}$~Mpc$^{-1}$ \cite{keyfinal}. 
However, if we assume that the calibration galaxies are internally reddened by 
$E(B-V) \sim 0.04$, then the zero point of the SBF system fades by 0.17~mag, 
and the SBF Hubble Constant increases to 75~km~s$^{-1}$~Mpc$^{-1}$.  This one 
correction is as large as the technique's entire systematic error budget.  
Such an error could not have been found without the cross-check provided by 
PNLF measurements.

\subsection{Comparisons with Measurements Outside the Distance Ladder}
No technique is perfectly calibrated, so distance measurements based on
secondary standard candles, such as the PNLF,  cannot avoid a component of
systematic uncertainty.  However, there are two galaxies in the local
universe with distance estimates that do not rely on the distance ladder.  The
first is NGC~4258, which has a resolved disk of cold gas orbiting its 
central black hole.  The proper motions and radial accelerations of water 
masers associated with this gas have been detected and measured:  the result 
is an unambiguous geometric distance to the galaxy of $7.2 \pm 0.3$~Mpc 
\cite{herrnstein}.  The second benchmark comes from the light echo of SN~1987A 
in the Large Magellanic Cloud.  Although the geometry of the light echo is 
still somewhat controversial, the most detailed and complete analysis of the 
object to date gives a distance of $D < 47.2 \pm 0.1$~kpc \cite{gould}.  In 
Table~1 we compare these values with the distances determined from the PNLF 
\cite{p12} and from the measurements of Cepheids \cite{keyfinal}.

\begin{table}
\caption{Benchmark Galaxy Distances}
\begin{center}
\setlength\tabcolsep{10pt}
\begin{tabular}{lccc}
\hline\noalign{\smallskip}
Method                &LMC    &NGC~4258 &$\Delta\mu$ (mag) \\
\hline
\noalign{\smallskip}
Geometry  &$< 18.37 \pm 0.04$ \phantom{$<$}
                                    &$29.29 \pm 0.09$  &$10.92 \pm 0.10$ \\
Cepheids  &18.50             &$29.44 \pm 0.07$  &$10.94 \pm 0.07$ \\
PNLF      &$18.47 \pm 0.11$  &$29.43 \pm 0.09$  &$10.96 \pm 0.14$ \\ 
\noalign{\smallskip}
\hline
\noalign{\smallskip}
\end{tabular}
\end{center}
\label{Tab1}
\end{table}
According to the table, the Cepheid and PNLF methods both overestimate
the distance to NGC~4258 by $\sim 0.14$~mag, \ie by $\sim 1.3 \, \sigma$ and 
$1.0 \, \sigma$, respectively.  In the absence of some systematic error 
affecting both methods, the probability of this happening is 
$\lesssim 5\%$.  On the other hand, there is no disagreement concerning
NGC~4258's distance {\it relative to that of the LMC:}  the Cepheids,
PNLF, and geometric techniques all agree to within $\pm 2\%$!  Such a
small error is probably fortuitous, but it does suggest the presence of
a systematic error in the entire extragalactic distance scale.

In fact, the {\sl HST Key Project\/} distances are all based on an LMC
distance modulus of $(m-M)_0 = 18.50$ \cite{keyfinal}, and, via the data
of Fig.~\ref{eps8}, the PNLF scale is tied to that of the Cepheids.  If the
zero point of the Cepheid scale were shifted to $(m-M)_0 = 18.37$, then all the 
measurements would be in agreement.  This consistency supports a shorter
distance to the LMC, and argues for a 7\% increase in the {\sl HST\/} Key
Project Hubble Constant to 77~km~s$^{-1}$~Mpc$^{-1}$.

\section{Future Directions}
The planetary nebula luminosity function is an excellent standard candle for 
measuring extragalactic distances within $\sim 20$~Mpc.  PNLF measurements are
precise, and, in terms of telescope time, much more efficient than variable
star monitoring or OB star spectroscopy.  However, the technique cannot be 
extended much farther.  Extragalactic PNe are point sources and their 
photometry is sky noise limited.  Hence to maintain a constant
signal-to-noise ratio, exposure times must grow as the fourth power of 
distance.  Since PNLF measurements in Virgo already require $\sim 4$~hr of 
4-m class telescope time in $\sim 1\arcsec$ seeing, observations at distances 
larger than $\sim 25$~Mpc are prohibitively expensive.  Improvements in seeing, 
telescope aperture, and instrumentation will help slightly, but the PNLF will 
never be competitive with techniques such as Surface Brightness Fluctuations 
or the Tully-Fisher relation.

On the other hand, PNLF observations are unlikely to disappear.  There will
always be some objects, such as NGC~4258, for which an additional, 
high-precision distance measurement is useful.  However, most future PNLF 
studies are likely to be performed as by-products of other investigations.
Planetary nebulae are powerful tools for the study of astrophysics and 
cosmology.  In addition to being excellent standard candles, PNe are useful 
probes of stellar population, unique tracers of chemical evolution, and 
excellent test particles for stellar kinematics and dark matter studies.
Moreover, photometry and spectroscopy of planetary nebulae is the best and 
perhaps only way to study the line-of-sight distribution and kinematics
of intracluster stars.   Our study of the evolutionary state of nearby galaxy
clusters has always been hampered by the limited number of test particles
available for study \cite{bing}.  However, these systems have plenty of
planetary nebulae -- in the core of Virgo alone, $\gtrsim 15,000$
intracluster planetaries are within reach of today's telescopes.  Thus
wide-field [O~III] $\lambda 5007$ imaging and follow-up spectroscopy in 
clusters such as Virgo and Fornax will be common in the coming decade.  

All these programs, from the study of chemical evolution to the analysis
of cluster kinematics, begin with the identification and photometric 
measurement of planetary nebulae.  PNLF distances will therefore continue to 
be measured in the local universe.

\null\par

I would like to thank G. Jacoby, J. Feldmeier, and P. Durrell for their
comments during the preparation of this paper.  This work made 
use of the NASA Extragalactic Database and was supported in part by NSF grant
AST 00-71238.

\null\par

%

\end{document}